\documentclass{elsart}
\usepackage{natbib}
\usepackage{graphicx}
\usepackage{amssymb}
\usepackage{amsmath}
\usepackage{bm}

\def\url#1{{\ttfamily\def\/{/\discretionary{}{}{}}#1}}
\def\bibcode#1{}

\begin{document}
\begin{frontmatter}
\title{Baryon oscillations}
\author{Martin White\thanksref{mjwemail}}
\address{Departments of Physics and Astronomy,\\
University of California, Berkeley, CA 94720}
\thanks[mjwemail]{mwhite@berkeley.edu}

\begin{abstract}
The coupling of photons and baryons by Thomson scattering in the early
universe imprints features in both the Cosmic Microwave Background (CMB)
and matter power spectra.  The former have been used to constrain a host
of cosmological parameters, the latter have the potential to strongly
constrain the expansion history of the universe and dark energy.
Key to this program is the means to localize the primordial features in
observations of galaxy spectra which necessarily involve galaxy bias,
non-linear evolution and redshift space distortions.  We present calculations,
based on mock catalogues produced from particle-mesh simulations, which
show the range of behaviors we might expect of galaxies in the real universe.
We find that non-linearity, galaxy bias and redshift space distortions
all introduce important modifications to the basic picture.
Both the galaxy bias and the (isotropic) redshift space distortions
lead to relatively smooth modulations in power on the scales of interest
to baryon oscillations.
The halo and galaxy power spectra exhibit low order structure beyond
the acoustic oscillations over the range of scales of relevance.
Fitting a cubic polynomial to the ratio of galaxy to dark matter power
reduces any remaining structure in the range
$0.03\le k\le 0.3\,h\,{\rm Mpc}^{-1}$ below the 2\% level,
which is close to the error in our calculations.
\end{abstract}
\end{frontmatter}

\section{Introduction}

Recently Eisenstein and collaborators \cite{SDSS}, using data from the
Sloan Digital Sky Survey\footnote{http://www.sdss.org/}, published evidence
for features in the matter power spectrum on scales of $100\,$Mpc.
These features, long predicted, hold the promise of another route to
understanding the expansion history of the universe and the influence
of dark energy.

It was realized very early \cite{GrandDad} that a universe with non-negligible
baryon fraction would result in significant features in the matter power
spectrum due to the coupling of the baryons to (hot) photons in the early
universe\footnote{In fact Sakharov (building on earlier work by Lifschitz)
predicted oscillations in the matter power spectrum even in a cold universe
with no radiation component -- the oscillations were an imprint of sound waves
as in the modern context, but the restoring pressure was caused by degenerate
electron pressure at high densities, not CMB photons.}.
When the universe was dense and highly ionized the baryons and photons were
tightly coupled by Thomson scattering.  During this phase the amplitude of
baryon-photon perturbations cannot grow, but rather undergo harmonic motion
with a slowly decaying amplitude.  Slightly after recombination the baryons
decouple from the radiation and the oscillations are frozen in -- leading to
subtle features in the total matter power spectrum analogous to the
(fractionally much larger) features in the power spectrum of the cosmic
microwave background (CMB) radiation.
The first calculations to emphasize the oscillations in the context of
hot big-bang CDM models were \cite{Early}.

\begin{figure}
\begin{center}
\resizebox{4.5in}{!}{\includegraphics{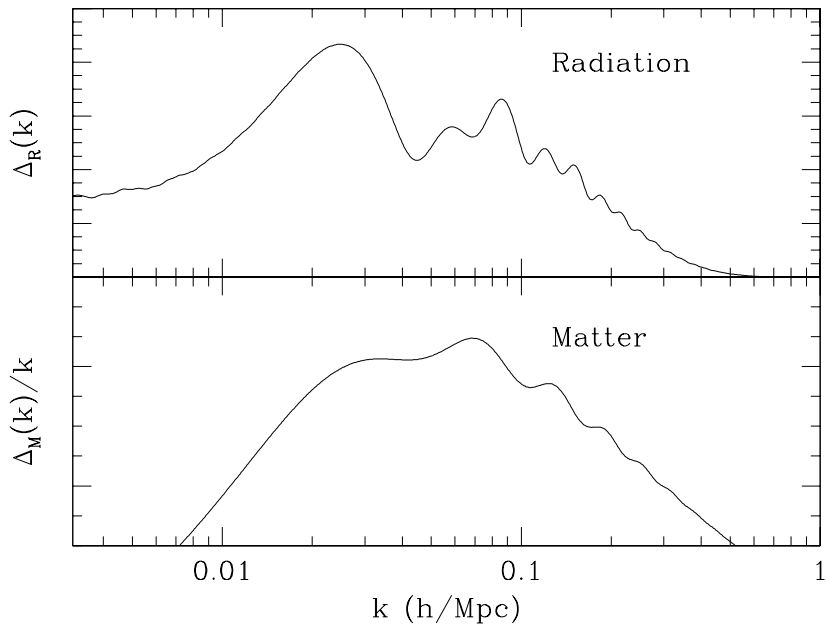}}
\end{center}
\caption{The rms fluctuations in the radiation and matter as a function
of wavenumber, $k$.  The structure in the CMB spectrum is familiar, in
its $\ell$-space form, from the WMAP measurements \protect\cite{WMAP}.
For the matter spectrum we have divided by $k$ to remove the overall trend
and enhance the appearance of the oscillations.  Note that the precise
relationship between the oscillations in the CMB and in the mass is complex
but the same physical processes affect both.}
\label{fig:rms}
\end{figure}

We show in Fig.~\ref{fig:rms} the matter and radiation power spectra,
in $k$-space, for a $\Lambda$CDM model (see below).  These spectra were
computed by numerical evolution of the coupled Einstein, fluid and Boltzmann
equations as described in \cite{Boltz,SSWZ}.
A description of the physics leading to the oscillations can be found
in \cite{EHSS} or Appendix A of \cite{MeiWhiPea}.
While the precise relationship between the positions of the peaks in
the radiation and matter power spectra is complex, the same physics
underlies both and they share a common scale: the sound horizon
\begin{equation}
  s = \int c_s(1+z)dt \simeq 147\pm 2\ {\rm Mpc}
\end{equation}
where $c_s$ is the sound speed and the measured value is taken from
\cite{WMAP}.  The peaks in the radiation spectrum occur roughly\footnote{These
relations are only approximate because of the expansion of the universe and
the evolving potentials and baryon to photon energy density ratio.} at
$ks=n\pi$ for integer $n$, the peaks in the matter spectrum are almost
out of phase.

It was pointed out in Refs.~\cite{CooHuHutJof,Eis03} that this scale could
be used as a standard ruler to constrain the distance-redshift relation, the
expansion of the universe and dark energy.  Numerous authors
\cite{Fisher} have now observed that a high-$z$ galaxy survey\footnote{It
is even possible that such oscillations could be seen in the Ly-$\alpha$
forest \cite{Davis} or in very large cluster surveys \cite{Ang}.} covering
upwards of several hundred square degrees could place interesting
constraints on dark energy.
Key to realizing this is the ability to accurately predict, from measurements
of the CMB, the physical scale at which the oscillations appear in the power
spectrum plus the means to localize those primordial features in observations
of galaxy spectra which necessarily involve galaxy bias, non-linear evolution
and redshift space distortions.
The former problem seems well in hand \cite{SSWZ,EisWhi}.
We make some preliminary investigations of the latter problem in this
paper.

\section{Fitting the extra physics} \label{sec:fitting}

The real space, linear theory matter power spectrum can be computed with
high accuracy \cite{SSWZ}, however we measure the non-linear galaxy power
spectrum in redshift space.
Our understanding of the transformation between these two is improving
rapidly, along with our ability to simulate the complex processes involved.
Thus to some degree the effects of non-linearity, bias and redshift space
distortions can be predicted.
However, we would like to reduce our dependence on detailed theoretical
calculations as much as possible.
One possibility is to parameterize the additional effects and measure the
parameters along with the sound horizon.  This amounts to marginalizing over
a range of models for the additional physics, inflating the error bars on
the cosmological parameters but reducing sensitivity to improper modeling
of non-linearity, bias or redshift space distortions.
In general we would like to allow enough freedom in our fit to account
for the effects while not degrading too severely our constraint on the
peak locations.  Improved theoretical modeling of non-linearity, bias and
redshift space distortions translate into stronger priors on the ``extra
physics'' parameters in this framework.

\begin{figure}
\begin{center}
\resizebox{4.5in}{!}{\includegraphics{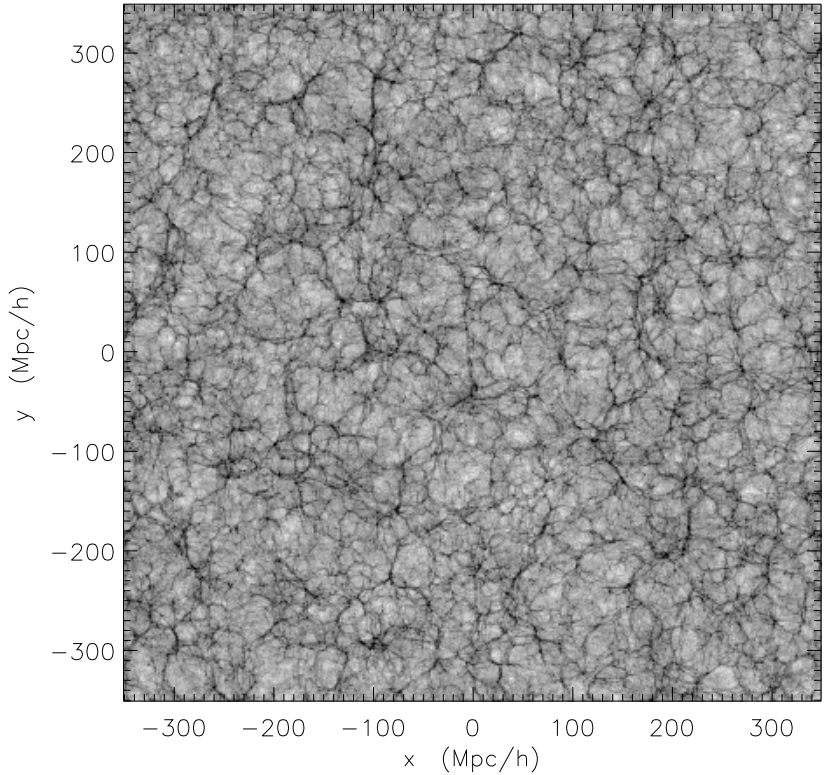}}
\end{center}
\caption{A slice, $10\,h^{-1}$Mpc thick, through the DM distribution in
the higher resolution run at $z=1$.  The greyscale is logarithmic in
density with black being $>10^2\bar{\rho}$ and white being
$<10^{-2}\bar{\rho}$.}
\label{fig:dmslice}
\end{figure}

Non-linearity, galaxy bias and redshift space distortions are all expected
to be ``simple'' on large scales.  Each has a characteristic scale
associated with it.
For non-linearity the scale is $k^{-1}_{\rm nl}\simeq 1-5\,h^{-1}$Mpc
for $z\simeq 0.3-3$ (see \S\ref{sec:results}).
For redshift space distortions the scale is set by the velocity dispersion
of the largest halos \cite{HaloRed}.  At $z\simeq 1$ we expect
$\sigma\simeq 500\,{\rm km}{\rm s}^{-1}\sim 5\,h^{-1}$Mpc.
Galaxy bias is expected to become simple on large scales \cite{SchWei}
and in halo-based models \cite{Halo} the transition is set by the virial
radius of the largest halo, $r_0\simeq 1\,h^{-1}$Mpc.
Thus it makes sense to fit, for $k<0.1\,h{\rm Mpc}^{-1}$,
\begin{equation}
  \Delta^2_{\rm obs}(k) = B^2(k) \Delta^2_{\rm dm}(k) + C(k)
\end{equation}
with $\Delta^2_{\rm dm}(k)$ the dimensionless power spectrum,
\begin{equation}
  \Delta^2(k) \equiv \frac{k^3P(k)}{2\pi^2}
  \quad {\rm with}\quad
  P(\vec{k})\equiv\left| \delta_{\vec{k}}\right|^2
\end{equation}
of the mass, in real space, and $B(k)$ and $C(k)$ are polynomials in $k$.
Within the halo model $B(k)$ describes the $2$-halo term while
$C(k)$ describes the $1$-halo term \cite{Nikhil}.
Since we are not computing $B(k)$ and $C(k)$ from a halo model formalism,
but merely exploring the relationship between $\Delta^2_{\rm obs}(k)$ and
$\Delta^2_{\rm dm}(k)$, we will take a slightly simpler approach (see later)
of absorbing $C(k)$ into a scale dependent bias, $b(k)$.
We note that the authors of Ref.~\cite{SeoEis05} make a different reduction
by taking $B(k)=$constant and absorbing any scale-dependence into $C(k)$.

As an example of the above approach we could expand the corrections to our
best fit theory in Hermite polynomials, $H_n$, which are bounded by
$-1\le H_n\le 1$.  Theoretical priors would then take the form of limits
on the size of the correction coefficients, for example $c_n$ is a Gaussian
of zero mean and dispersion $\sigma_n$ which parameterizes the residual
uncertainty in our modeling.
In this paper we make a preliminary investigation of the structure of
these functions using numerical simulations.

\section{Simulations} \label{sec:sim}

The basis for our calculations is a sequence of particle-mesh (PM)
simulations of a $\Lambda$CDM cosmology ($\Omega_M=0.28=1-\Omega_\Lambda$,
$\Omega_B\simeq 0.049$, $n=1$, $\sigma_8=0.9$ and $h=0.7$)
in a periodic, cubical box of side $700\,h^{-1}$Mpc=$1\,$Gpc.
The main simulation evolved $1024^3$ particles of mass
$2.5\times 10^{10}\,h^{-1}M_\odot$ from $z=60$ to $z=0$ with a regular
mesh of $2048^3$ points used to compute the forces.
Outputs at both fixed time and along a light-``cone'' were produced at
$z=4$, 3, 2, 1.5, 1, 0.3 and 0 (see Fig.~\ref{fig:dmslice}).
To make the geometry easier for analysis we assumed the infinitely
distant observer approximation in producing the light-cone outputs.
The rays did not converge on the origin, but rather were all parallel
to the $\hat{z}$-axis of the simulation box, respecting the periodic
nature of the simulation.
For some of our results on the rarer, more massive halos we used a
series of 9 lower resolution ($512^3$) simulations with the same size box.

For the initial conditions we used the linear theory power spectrum from
Fig.~\ref{fig:rms} and displaced particles from a ``fuzzy'' regular grid
using the Zel'dovich approximation.
For the main simulation we picked random phases, but for the amplitudes
of the low-$k$ modes we chose an abnormally common set.
The power spectrum at low-$k$ thus shows significantly less scatter than a
Rayleigh distribution, making it easier to see the baryonic features.
Moving to higher $k$, with more modes in the box, we gradually moved to a
Rayleigh distribution so that the initial conditions properly sampled a
Gaussian distribution.
Since we are primarily interested in the effects of galaxy modeling,
redshift space distortions and light-cone evolution the exact initial
conditions are merely a convenience.
For the smaller runs we chose unconstrained Gaussian initial conditions,
since we can average over several realizations.

For each output we generate a catalog of halos using the Friends-of-Friends
algorithm \cite{FoF} with a linking length of $b=0.15$ in units of the
mean inter-particle spacing.
This procedure partitions the particles into equivalence classes, by linking
together all particle pairs separated by less than a distance $b$.
The groups correspond roughly to all particles above a density of
$3/(2\pi b^3)\simeq 140$ times the background density and we keep all
groups with more than 20 particles.

\begin{figure}
\begin{center}
\resizebox{4.5in}{!}{\includegraphics{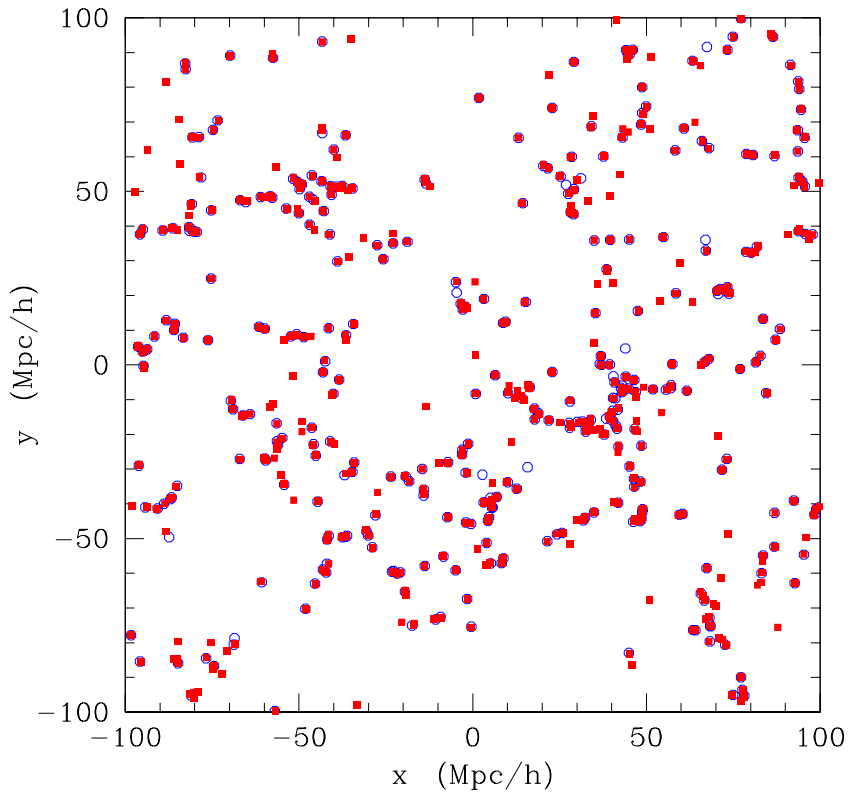}}
\end{center}
\caption{The positions of halos hosting galaxies in a $10\,h^{-1}$Mpc
slice through a $200\,h^{-1}$Mpc simulation of a $\Lambda$CDM model with
a high force resolution code \protect\cite{TreePM} and the PM code used
in this work.  The simulations each evolved $256^3$ particles from the
same initial conditions to $z=0$.  The filled (red) squares are halos from
the high-resolution run, the (blue) open circles those from the PM run.}
\label{fig:grppos}
\end{figure}

\begin{figure}
\begin{center}
\resizebox{4.5in}{!}{\includegraphics{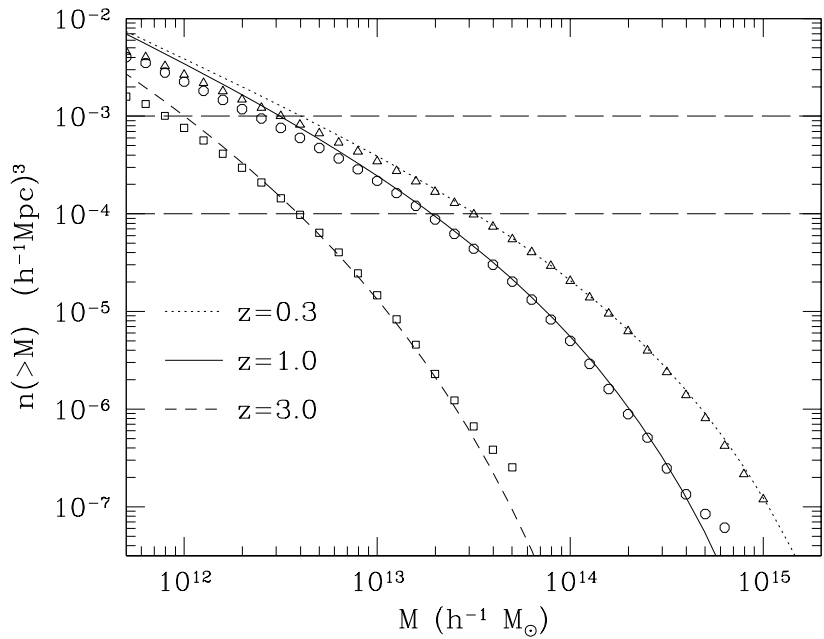}}
\end{center}
\caption{The mass function of halos in the PM simulation at $z=3$, 1 and 0.3.
The lines indicate the prediction of Ref.~\protect\cite{ST}.  The shortfall
at low mass is due to the limited force resolution of the PM simulation.}
\label{fig:massfn}
\end{figure}

The huge advantage of a PM simulation over one involving higher force
resolution is its speed or low cost.  The current simulations, plus all of
the post-processing took only $\sim 10^4$ CPU hours on the IBM-SP Seaborg at
NERSC --- a few percent of the resources required for the Hubble volume
\cite{HV} or Millennium runs \cite{Millenium}.  It is easy to conceive of
running large numbers of PM simulations of this size to explore parameter
space or provide statistical samples -- but is a PM simulation adequate?

We would like to follow a large number of particles to accurately evolve
the modes in the quasi-linear regime, but the baryon oscillation scales are
very large so don't require a huge dynamic range.
Indeed PM simulations on a 1990's era mid-level desktop workstation were
perfectly adequate to resolve the relevant gravitational physics and explore
simple models of galaxy bias \cite{MeiWhiPea}.
Our main simulation resolves the relevant scales with hundreds of mesh
cells in each dimension.

The issue of galaxy modeling is more subtle.  Clearly our simulations are
not correctly resolving the internal structure of dark matter halos hosting
galaxies, and are completely unable to resolve sub-structure.  This makes
careful modeling of galaxy formation impossible.
We shall take a different approach however.  Rather than trying to
model as accurately as possible the physics of galaxy formation, we shall
try to get a feel for the range of possibilities afforded by models in
which galaxies form in dark matter halos.  If we are willing to make
assumptions about the spatial and velocity distribution of galaxies in
halos and their halo occupation distribution then we need only resolve
the positions and masses of the hosting halos in the simulation in order
to make realistic mock catalogs -- an approach similar to that of
Ref.~\cite{PTHalos}.

Fortunately the masses and positions of the halos are reasonably well
modeled, as we show in Fig.~\ref{fig:grppos}.  The figure shows the $x$
and $y$ positions of halos which would host galaxies in a thin slice through
two simulations.  Both simulations have the same initial conditions, one is
run with a high force resolution code \cite{TreePM} and one with the PM
code\footnote{The force resolution of the main, $1024^3$, run in this paper
is in fact slightly higher than that shown in the figure -- which was done
before the large run was started.}.
The overall agreement in positions and masses is very good, although
there is a tendency for the PM simulation to merge some close halos that the
higher resolution simulation resolves as discrete.  A few of the halos
seen in one simulation but not the other have their centers falling just
outside of the thin slice in one run but inside in the other.
The mass function of the simulation at $z=3$, 1 and 0.3 is shown in
Fig.~\ref{fig:massfn} compared to the fitting function of Ref.~\cite{ST}.
There is a shortfall at lower masses and an excess at the very rare,
high mass end -- the excess appears to come from the initial conditions
used, the shortfall at low mass comes from the finite force resolution.
For halos in the most interesting range for us,
few $\times 10^{12}$ -- $10^{14}\,h^{-1}M_\odot$, the
agreement is satisfactory.
Based on these comparisons we believe that our simulations, while
inappropriate to model small scale clustering, are adequate for our purposes.

\begin{table}
\begin{center}
\begin{tabular}{cccc}
$\log_{10}M_{\rm min}$ & $\log_{10}M_1$ & $\langle b\rangle$ \\
  12.8                 & 13.0           & 2.4 \\
  12.6                 & 13.5           & 2.0 \\
  12.5                 & 14.0           & 1.8 \\
  12.5                 & 14.5           & 1.8
\end{tabular}
\end{center}
\caption{The HOD parameters, for the form of Eq.~\protect\ref{eqn:hod},
we have used for some of our models at $z=1$.  Masses are in units of
$h^{-1}M_\odot$ and the number density for all models is
$10^{-3}\,h^3\,{\rm Mpc}^{-3}$.}
\label{tab:hod}
\end{table}

To make the mock galaxy samples we then take an output and choose a mean
occupancy of halos:
$N(M)\equiv\left\langle N_{\rm gal}(M_{\rm halo})\right\rangle$.
Each halo either hosts a central galaxy or does not.
For each halo we define a galaxy to live at the minimum of the halo
potential, with the center of mass velocity, with probability
$p={\rm max}[1,N(M)]$.  Following Ref.~\cite{KBWKGAP}, if $N(M)>1$ the
mean number of satellites, $N_{\rm sat}=N(M)-1$, is computed for the
halo and a Poisson random number, $n_{\rm sat}$, drawn.
Then $n_{\rm sat}$ dark matter particles, chosen at random but with weights
depending on their radial distance from the potential minimum, are anointed
as galaxies.  Our fiducial model has equal weights -- the satellite galaxies
trace the dark matter.  The galaxy velocity is taken to be
$v_{\rm sat} = v_{\rm com} + A \left(v_{\rm part}-v_{\rm com}\right)$ where
$v_{\rm part}$ is the particle velocity and $v_{\rm com}$ is the velocity of
the halo center-of-mass and our fiducial model has $A=1$,
i.e.~$v_{\rm sat}=v_{\rm part}$.
The result is a set of ``galaxies'' whose clustering properties can
be computed.  While the galaxy model is not prescriptive, or likely even
close to ``right'', it is physically well motivated, easy to adjust and
leads to galaxy catalogs with non-linear, scale-dependent, stochastic (and
in principle luminosity dependent) biasing, redshift space distortions and
(later) light-cone effects.

\begin{figure}
\begin{center}
\resizebox{4.5in}{!}{\includegraphics{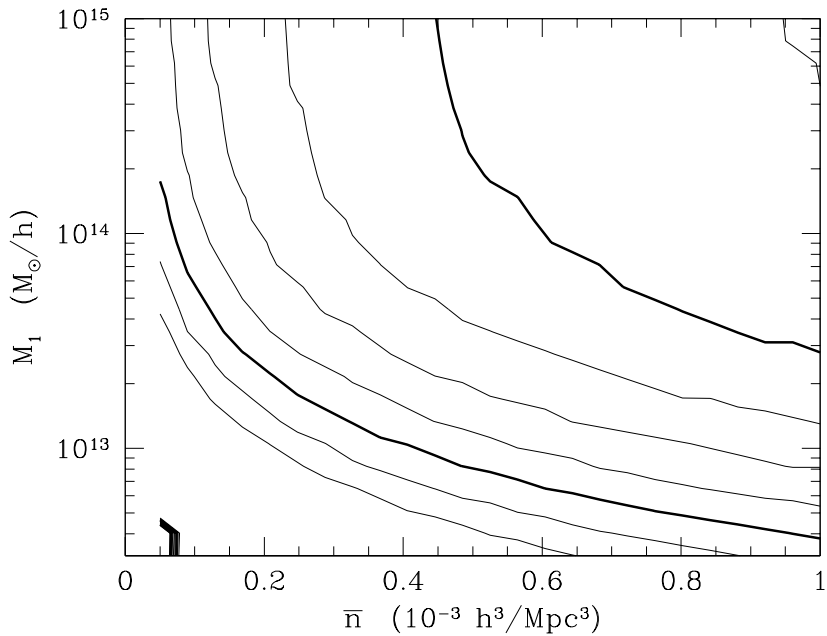}}
\end{center}
\caption{The large-scale bias at $z=1$ as a function of number density,
$\bar{n}$, and $M_1$ for a HOD of the form of Eq.~\protect\ref{eqn:hod}.
Contours are spaced in units of 0.25.  The bias increases to lower left
and $\langle b\rangle=2$ and 3 are shown as thick contours.}
\label{fig:bias}
\end{figure}

Even with the distribution of galaxies chosen there is still tremendous
freedom in specifying $N(M)$.  For this initial exploration we choose a
very simple two-parameter form with
\begin{equation}
  N(M) = \Theta(M-M_{\rm min})\ \frac{\left(M-M_{\rm min}\right)+M_1}{M_1}
\label{eqn:hod}
\end{equation}
where $\Theta(x)$ is the Heavyside step function.
If we take $M_1\to\infty$ our catalog reduces to a catalog of halos
more massive than $M_{\rm min}$.  By holding $\bar{n}$ fixed we can
specify a 1-parameter sequence of models with varying $M_{\rm min}$
or large-scale bias, $\langle b\rangle$, as shown in
Fig.~\ref{fig:bias}.  A selection of models is given in
Table \ref{tab:hod}.

\begin{figure}
\begin{center}
\resizebox{4.5in}{!}{\includegraphics{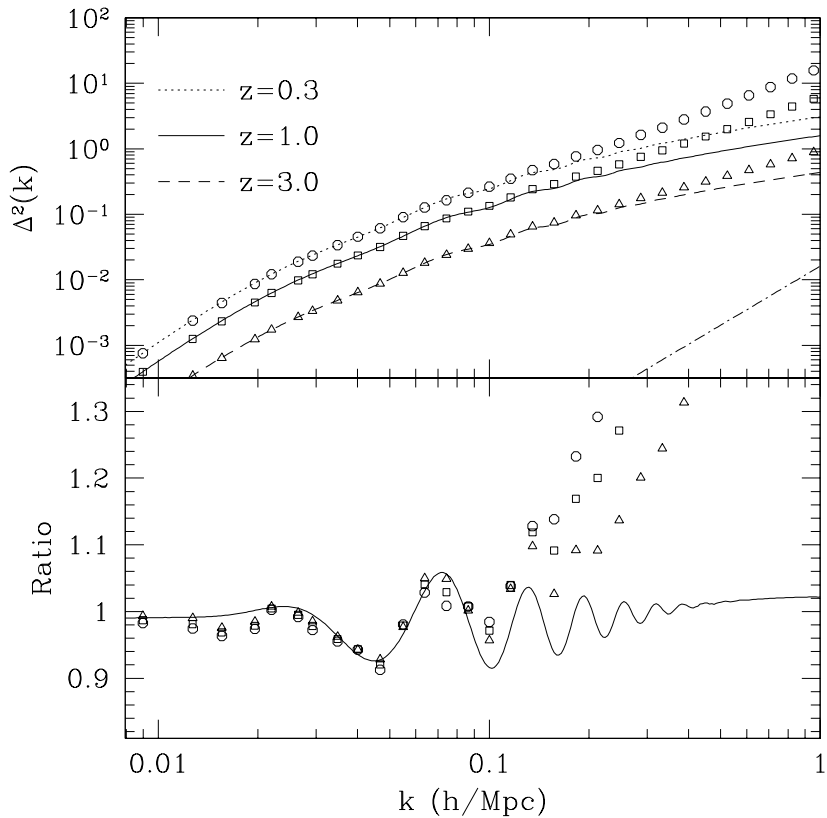}}
\end{center}
\caption{The real space matter power spectrum from 3 constant time slices.
In the upper panel the dimensionless power spectrum (points) is shown at
$z=0.3$, 1 and 3 along with the linear theory prediction (lines).
Shot-noise is not subtracted.  The shot-noise contribution, for Poisson
sampling, is shown by the dot-dashed line in the lower right.  It
corresponds to $\bar{n}\simeq 3\,h^3\,{\rm Mpc}^{-3}$.
The lower panel plots the ratio of the results to a ``smooth'' spectrum --
the no-baryon fit of Ref.~\protect\cite{EisHu} -- with the linear growth
factored out.}
\label{fig:pkmm}
\end{figure}

For most of our results we will concentrate on the $z=1$ output.
We still do not know what the best sample of galaxies is for detecting
baryon oscillations at $z=1$.  Since it is not our intention to model
a particular strategy we will choose $\bar{n}=10^{-3}\,h^3\,{\rm Mpc}^{-3}$
-- at the upper end of the range advocated by \cite{SeoEis03} for $z=1$.
We find for samples with lower number density the shot-noise correction
is very large\footnote{The shot-noise power scales as $\bar{n}^{-1}$.  The
dot dashed line in Fig.~\ref{fig:pkmm} shows the value for
$\bar{n}\simeq 3\,h^3\,{\rm Mpc}^{-3}$.} and the sampling noise unacceptably
high for our limited volume.
A density of $10^{-4}$ is close to the observed number density of the SDSS
luminous red galaxy (LRG) sample with $\langle z\rangle\simeq 0.3$.
These galaxies have a large-scale bias of approximately two \cite{Zehavi}.
In our model, at this redshift, we can fit the number density and bias with
$M_{\rm min}\simeq 3.3\times 10^{13}\,h^{-1}M_\odot$ and
$M_1\simeq 10^{15}\,h^{-1}M_\odot$.  If we imagine the HOD doesn't evolve
with redshift, and with $M_1$ this large, the LRGs are essentially the
central galaxy population of halos above $M_{\rm min}$.  Since this mass
is quite large, and the halos quite rare, we use the lower resolution runs
to study such massive halos.
In future, with more simulations in hand, we could enlarge our study to
other populations.
Finally we note that for all of these number densities the contribution to
the power spectrum from shot-noise, assuming it is Poisson, is as large as
the signal on scales of $k\simeq 0.1\,h{\rm Mpc}^{-1}$.
This is no accident -- in the Gauss-Poisson limit the optimal survey
strategy for constraining the power spectrum is to balance the cosmic
variance and shot-noise power \cite{SparseSampling}.
Thus shot-noise subtraction is an important part of the power spectrum
calculation and deserves further study.

\begin{figure}
\begin{center}
\resizebox{4.5in}{!}{\includegraphics{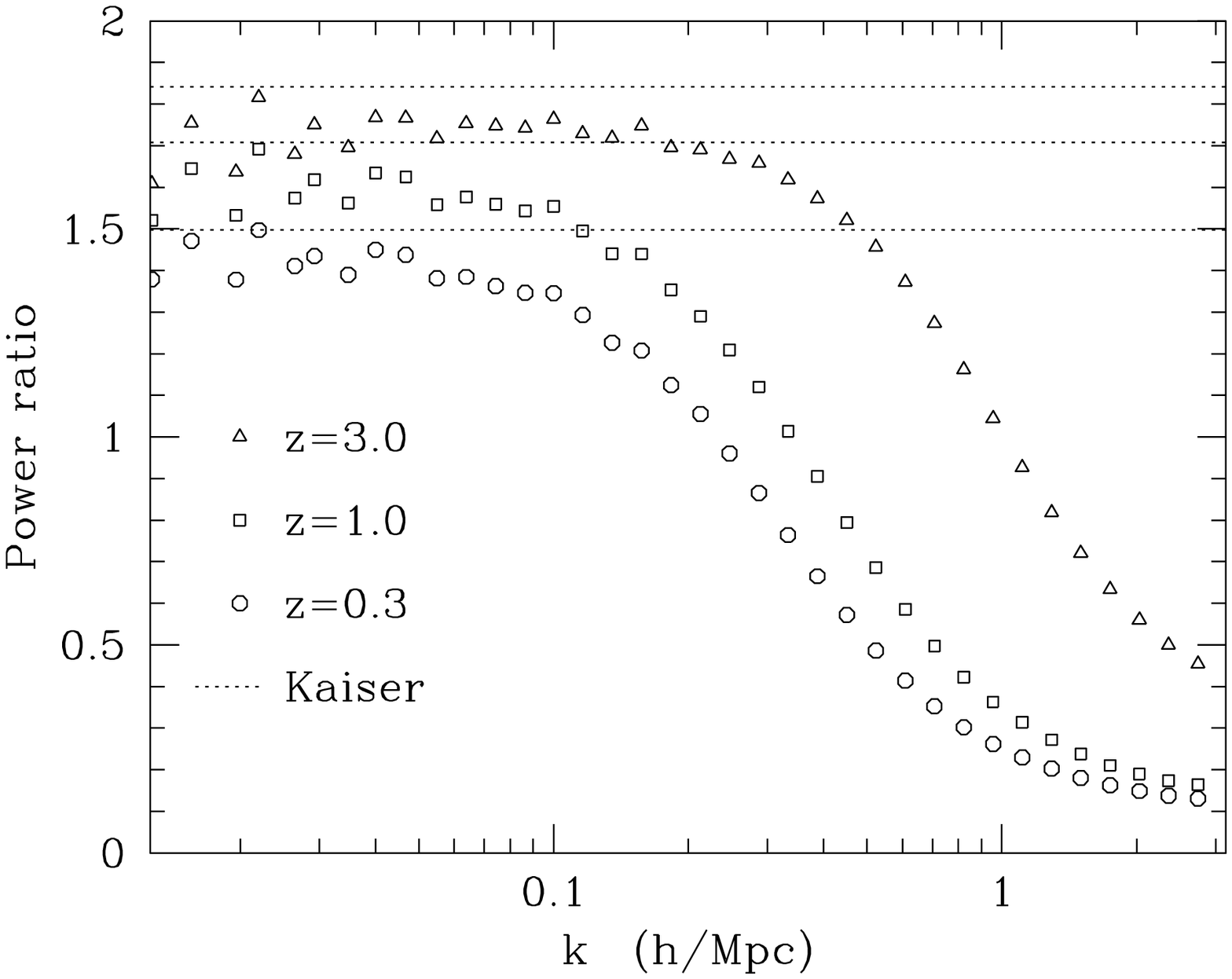}}
\end{center}
\caption{The ratio of the (angle averaged, dark matter) redshift space
power spectrum to the real space power spectrum at $z=0.3$, 1 and 3.
The horizontal dotted lines are the prediction of linear theory
\protect\cite{Kaiser} -- $1+\frac{2}{3}\beta+\frac{1}{5}\beta^2$ with
$\beta\simeq\Omega_m^{0.6}$ -- which should be valid as $k\to 0$.
The points are highly correlated.}
\label{fig:red}
\end{figure}

\section{Results} \label{sec:results}

We begin by looking at the evolution of the dark matter.  The dimensionless
power spectrum of the mass, in real space, is shown in Fig.~\ref{fig:pkmm}
along with the linear theory predictions.
The power spectrum was computed by assigning the particles to the nearest
grid point of a regular, $1024^3$ Cartesian mesh and Fourier transforming
the resultant density field.  The resulting $P(k)$, corrected for the
assignment to the grid using the appropriate window function, were placed
in logarithmically spaced bins of $k$.  The average $P(k)$ is plotted at
the position of the average $k$ in each bin.

Fig.~\ref{fig:pkmm} shows the effects of non-linear evolution on the power
spectrum.  The trends are familiar \cite{MeiWhiPea,Millenium,Coo04}: first
a decrease in power, compared to linear theory, in the trans-linear regime
followed by an increase in the non-linear regime.
It is clear from Fig.~\ref{fig:pkmm} that the mode coupling induced by
non-linear evolution erases the oscillations at high-$k$ \cite{MeiWhiPea},
with the feature just beyond $k\simeq 0.1\,h\,{\rm Mpc}^{-1}$ being
unavailable at low-$z$.
The scale of non-linearity, where $\Delta^2(k_{\rm nl})=1$, is only an
approximate guide to the redshift dependence of the mode coupling because
of the increasing flatness of $\Delta^2(k)$ as we push to higher $k$.
Even though one can still see remnants of the oscillatory features
at $0.2\,h\,{\rm Mpc}^{-1}$, the $k$-modes are no longer independent
\cite{MeiWhi,CooHu}.
An accurate calculation of the covariance matrix requires more simulations
than we have in hand \cite{MeiWhi}.
Given its importance for parameter forecasts however we make an estimate
{}from our 9 lower resolution runs -- for the $z=1$ mass power, neighboring
bins are correlated at the many tens of percent level beyond
$k\simeq 0.1\,h\,{\rm Mpc}^{-1}$ increasing to above 90\% for
$k\simeq 0.2\,h\,{\rm Mpc}^{-1}$.  Further mapping of these correlations
is left for future work.

Fig.~\ref{fig:red} shows the effect of redshift space distortions on the
angle-averaged power spectrum.  The power is enhanced, by a $z$-dependent
factor, on large scales due to supercluster infall \cite{Kaiser} and
suppressed on small scales due to virial motions within halos.
The power spectrum in redshift space was computed assuming the distant
observer approximation for all outputs and the periodicity of the simulation
was used to remap positions.
The power ratios do not recover to the results of Ref.~\cite{Kaiser} on
large scales.
A test with a different random number choice for the initial conditions
came closer for the lowest $k$ modes, indicating that sampling variance
accounts for some of the disagreement.
Whether we should reach the limits shown on the scales relevant to baryon
oscillations remains in doubt -- see Ref.~\cite{Sco04} and references
therein for further discussion.

\begin{figure}
\begin{center}
\resizebox{4.5in}{!}{\includegraphics{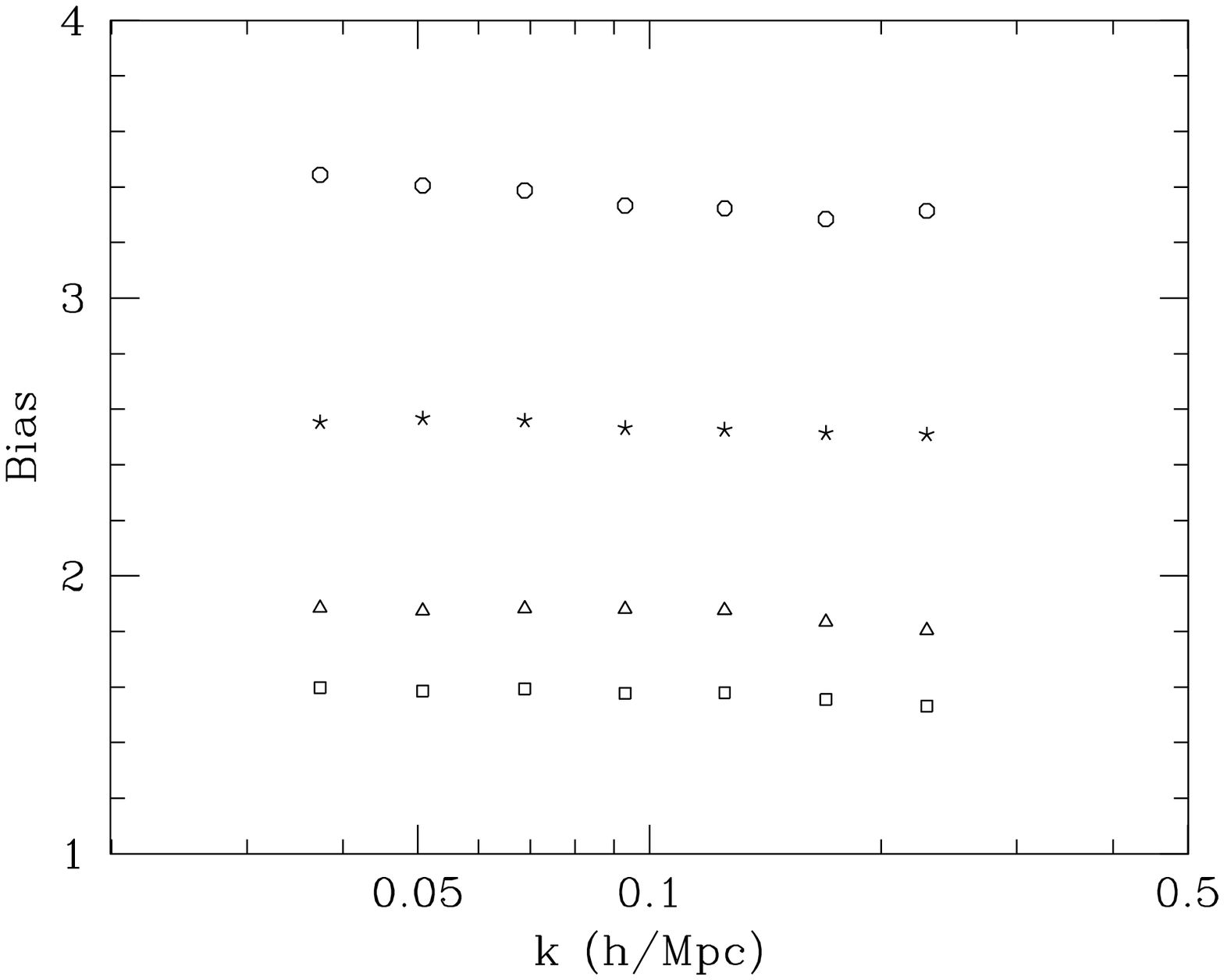}}
\resizebox{4.5in}{!}{\includegraphics{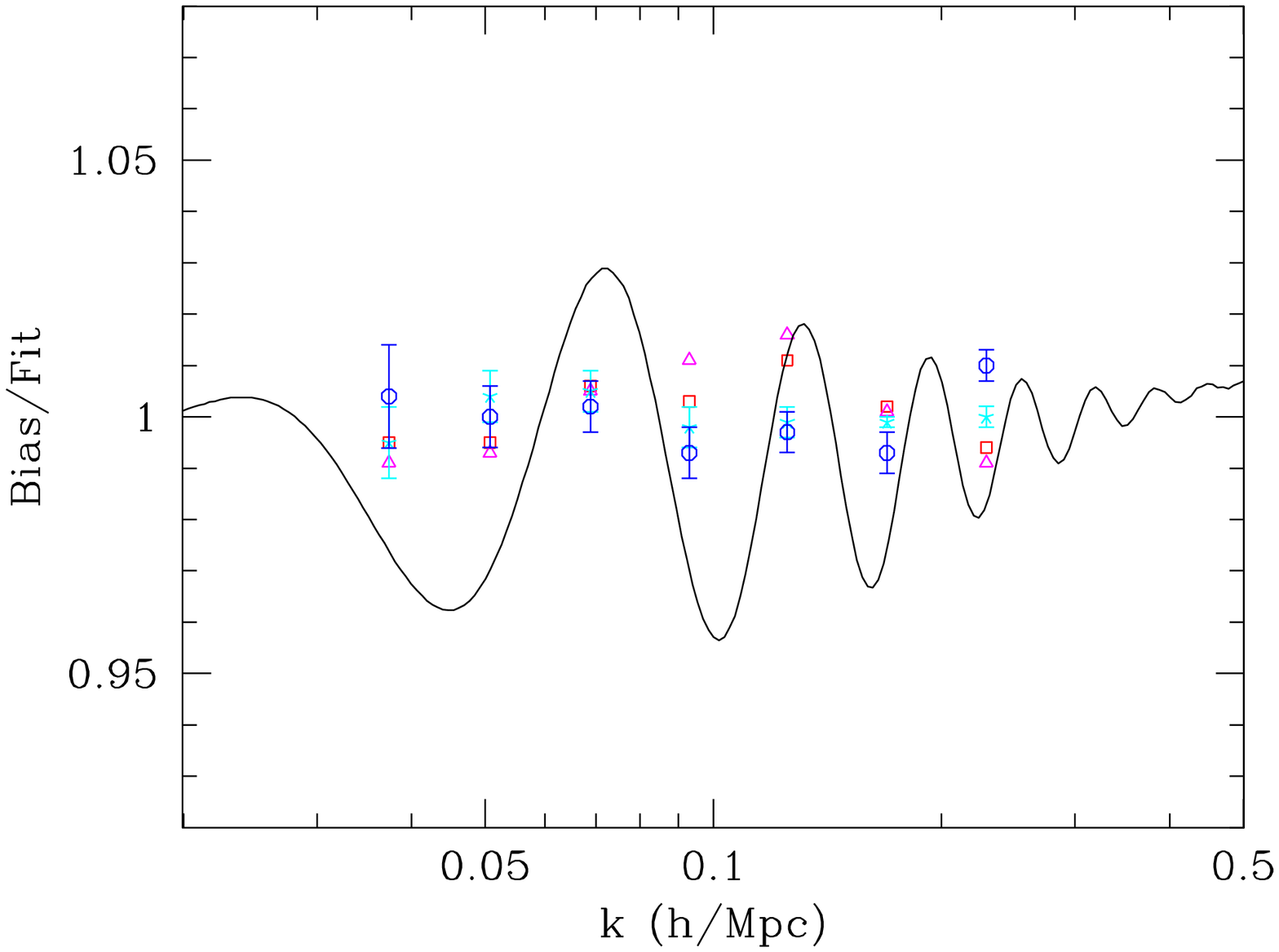}}
\end{center}
\caption{Top: The halo bias, defined as in Eq.~\protect\ref{eqn:bk},
for halos above $M_{\rm min}=10^{12}$ (squares), $10^{12.5}$ (triangles),
$10^{13}$ (stars) and $10^{13.5}\,h^{-1}M_\odot$ (circles) at $z=1$.  We
have subtracted shot-noise from the halo power spectra, assuming it is
Poisson.  The points are highly correlated.  Bottom: The halo bias, divided
by a best fit line, in the interval $0.03<k<0.3\,h{\rm Mpc}^{-1}$, compared
to the oscillation signal.}
\label{fig:halobias}
\end{figure}

Now we turn to the halo and galaxy catalogues, first in real space.
Fig.~\ref{fig:halobias} shows the scale-dependent bias of halos more
massive than $M_{\rm min}$ at $z=1$.  Here bias is defined as
\begin{equation}
  \Delta^2_{\rm halo}(k)=b^2(k)\Delta^2_{\rm dm}(k)
\label{eqn:bk}
\end{equation}
where $\Delta^2_{\rm dm}$ is the non-linear dark matter spectrum -- the
bias thus contains a mix of non-linear and galaxy formation physics.
Since the features we are looking for are small, we fit a linear trend
(in $\ln k$) to each set of points over the range
$k=0.03-0.3\,h{\rm Mpc}^{-1}$ and plot the ratio of the bias to the fit
in the lower panel of Fig.~\ref{fig:halobias}.
What we would like to see is the points being essentially unity.

For the most numerous halos a jackknife estimate suggests that the error
bars are smaller than 3\% on scales $k>0.03\,h{\rm Mpc}^{-1}$, though the
errors are correlated.  One must be careful of jackknife estimates on
scales approaching the box size, but nonetheless this implies that any
structure is smaller than the signal near $k\simeq 0.07\,h{\rm Mpc}^{-1}$
(but may influence the peak positions at the accuracy with which they
need to be determined to constrain dark energy).
Unfortunately the errors increase as we go to the rarer halos, and for
$M_{\rm min}=10^{13.5}\,h^{-1}M_\odot$ the errors are 2-7\% over the
range $k=0.03-0.3\,h{\rm Mpc}^{-1}$.
For this reason we switch to the lower resolution runs for the halos
with $M_{\rm min}\ge 10^{13}\,h^{-1}M_\odot$ and use the dispersion
between simulations to estimate the errors.

\begin{figure}
\begin{center}
\resizebox{4.5in}{!}{\includegraphics{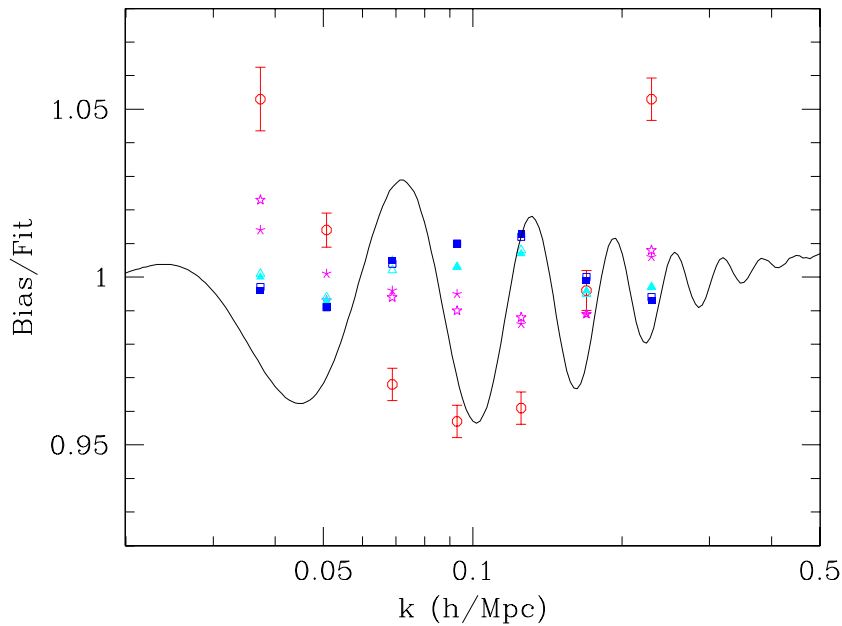}}
\end{center}
\caption{The galaxy bias, divided by a best fit line, in the interval
$0.03<k<0.3\,h{\rm Mpc}^{-1}$, compared to the oscillation signal.
The fit is linear in $\ln(k)$ and weights each point equally.
Models with $\bar{n}=10^{-3}\,h^{3}{\rm Mpc}^{-3}$ are shown with
$\log_{10}M_1=13.0$ (circles), $13.5$ (stars), $14.0$ (triangles)
and $14.5$ (squares).  Two realizations for each of the last 3 models are
plotted.  For the model with $\log_{10}M_1=13$ we use the average of
models from the 9 lower resolution runs, plotting the error in the mean
rather than multiple realizations.}
\label{fig:relbias_g}
\end{figure}

We see in Fig.~\ref{fig:halobias} some curvature remaining after the
linear trend is removed.  The size of the curvature is small, and could
be simply noise from the finite number of modes simulated.
It is also possible that this reflects a modulation in the bias of rare halos
(which in the peaks formalism \cite{PeaksBias} depends strongly on the
linear mass variance) when compared to the underlying non-linear power of
the dark matter.  In either case the modulation appears to be smooth, and
of lower amplitude than the baryon signal.

Next we show results for some of our galaxy catalogues, with
$\bar{n}=10^{-3}\,h^3\,{\rm Mpc}^{-3}$.
The trend-removed bias is shown in Fig.~\ref{fig:relbias_g}.
At the level required for baryon oscillations the bias is clearly not
constant, exhibiting curvature once the best-fit line is removed.
A comparison of catalogues produced with different random number seeds
gives some sense of the error, but only enters in the positions of the
satellite galaxies.  A jackknife estimate of the errors suggests they
are less than 2\% in the range $k=0.03-0.3\,h{\rm Mpc}^{-1}$ for all
of the models, again with the usual caveat about jackknife errors.
For the models with the largest $M_{\rm min}$, which show the biggest
curvature, we can use the 9 lower resolution runs to estimate the significance
of the curvature.  The agreement is fairly good between the average of the 9
low resolution runs and our fiducial model, with the former showing an error
around 1\% over the $k$-range shown.

\begin{figure}
\begin{center}
\resizebox{4.5in}{!}{\includegraphics{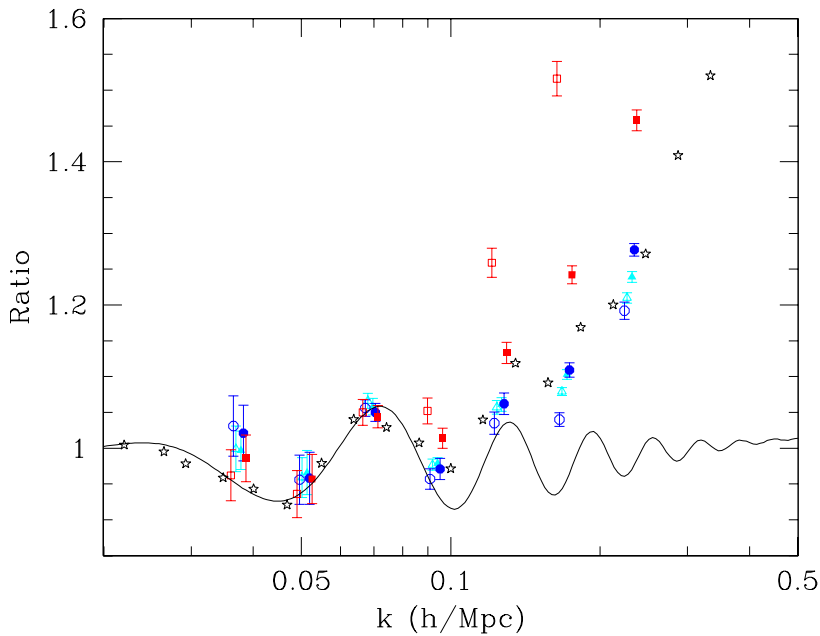}}
\end{center}
\caption{The real (open) and redshift (filled) space, halo power spectrum
at $z=1$, divided by the featureless linear theory prediction of
Ref.~\protect\cite{EisHu} and a constant bias.  We show results for the
halos most likely to host LRGs, $M_{\rm min}=10^{13}$ (triangles) and
$10^{13.5}\,h^{-1}M_\odot$ (circles), and for our HOD with
$M_1=10^{13}\,h^{-1}M_\odot$ (squares).  The points have been slightly
offset (horizontally) for clarity.
The stars are the non-linear mass power spectrum from
Fig.~\protect\ref{fig:pkmm}.
The solid line is again the linear theory prediction from numerical
integration of the coupled Einstein, Boltzmann and fluid equations.}
\label{fig:postfit}
\end{figure}

Now we shift to redshift space.
The low-$k$ redshift space enhancements for our galaxy and halo catalogues
are (fractionally) smaller than for the matter, due to the positive bias of
the tracers \cite{Kaiser}.  The models with a satellite galaxy component
show a suppression at high-$k$ that begins slightly before that in the matter
while the centers of halos show much less suppression, with the real and
redshift space power comparable at high-$k$.
The latter is due to our assumption that the central halo velocity is the
center of mass velocity of the halo.  This suppresses the virial motions
for these points, and it is the virial motion which leads to the depletion
of small-scale power.  The actual velocity of central galaxies in halos
can be determined observationally from their power spectrum, and is likely
to be higher than our center of mass model assumes.
Since the power suppression is relatively smooth our approximation will not
alter our basic conclusions.

In principle there is more information in the redshift space power
spectrum than shown in Fig.~\ref{fig:red} because it is no longer
isotropic.  We shall not investigate this degree of freedom in this
paper.   Though in principle the information is included in the simulations,
our computational volume is too small to allow multiple sub-divisions of
the sample while maintaining reasonable significance.  
For the angle-averaged power spectrum it is clear that on the scales
relevant to the oscillations redshift space distortions are a very
slowly varying function for $z\ge 1$.

Fig.~\ref{fig:postfit} shows a slightly different view of these results.
Here we show the redshift space, galaxy power spectrum at $z=1$ divided
by the featureless linear theory prediction of Ref.~\cite{EisHu} and a
constant bias fit to the points
$0.03\le k\le 0.1\,h\,{\rm Mpc}^{-1}$.
Superposed is the linear theory prediction from numerical integration of
the coupled Einstein, Boltzmann and fluid equations \cite{Boltz,SSWZ}.
The degree to which these differ shows the level of correction that
needs to be applied.

{}From these studies it appears that a cubic fit (in $\ln k$) to the
bias is sufficient to remove the scale dependence of the bias to the
percent level in the range $k=0.03-0.3\,h{\rm Mpc}^{-1}$.  This deviation
is approximately the same size as the error in our calculation.
To quantify this better would require more or larger simulations.

\section{Conclusions} \label{sec:conclusions}

The coupling of baryons and photons by Thomson scattering in the early
universe leads to a rich structure in the power spectra of the CMB
photons and the matter.  The study of the former has revolutionized
cosmology and allowed precise measurement of a host of important
cosmological parameters.  The study of the latter is still in its
infancy, but holds the potential to constrain the nature of the dark
energy believed to be causing the accelerated expansion of the universe.

Key to the baryon oscillation program is the ability to compare future
observations to theoretical computations of the features in the matter
power spectrum.  Such calculations are under exquisite control in linear
theory \cite{SSWZ} and the next stage is to explore the effects of
non-linear evolution, galaxy biasing and redshift space distortions.
Non-linear evolution can be straightforwardly modeled with the current
generation of N-body simulations, so the main difficulty lies in
understanding or parameterizing galaxy formation.

There have been tremendous advances in our ability to model and understand
the relation between galaxies and halos in the last few years.
The earlier generation of biasing models \cite{MeiWhiPea} were highly
simplistic, and today we can do significantly better using halo based
schemes \cite{Halo}.  Indeed results based on sophisticated modeling of
galaxies within halos \cite{Millenium} or in the limit where the halos
are the tracers \cite{Ang} have already been published.  Modeling along
similar lines to that done here has recently been reported in
Ref.~\cite{SeoEis05}.
Where this is overlap our results are in good agreement.

In this paper we explored a number of phenomenological models for
populating halos in PM simulations with galaxies.  While unlikely to
be correct in any detail these schemes serve to delineate the range
of behaviors we might expect in the real universe.
We find that non-linearity, galaxy bias and redshift space distortions
all introduce important modifications to the basic picture.
Both the galaxy bias and the (isotropic) redshift space distortions
lead to relatively smooth modulations in power on the scales of interest
to baryon oscillations, at least to the level of accuracy we are able
to achieve here.
The halo and galaxy power spectra exhibit low order structure beyond
the acoustic oscillations over the range of scales of relevance.
Fitting a cubic polynomial to the ratio of galaxy to dark matter power
reduces any remaining structure in the range
$0.03\le k\le 0.3\,h\,{\rm Mpc}^{-1}$ below the 2\% level,
comparable to the uncertainty in our calculations.
For future, ambitious, surveys even this level of uncertainty may
represent a significant piece of the error budget on dark energy properties.
Redshift space distortions remain to be explored in more detail.
The relative velocity of central galaxies and halos affects the redshift
space power spectrum of the former on interesting scales, and has not
been well modeled here.  In addition there is useful information in the
angular structure of the redshift space distortions which we do not have
the statistics to explore carefully.
We hope to return to some of these issues in a future publication.

Future large redshift surveys offer the opportunity to measure a
characteristic scale in the universe: the sound horizon at the time
of photon-baryon decoupling.  This standard ruler, which we can calibrate
{}from observations of the CMB, may allow us to tightly constrain the
evolution of the scale factor and determine the nature of dark energy.
To ensure the success of these efforts we need to improve our understanding
of the theoretical underpinnings of the method and generate simulated
universes which can be used to refine and test our observational strategies.

I would like to thank N. Padmanabhan for conversations and collaboration
on marginalizing out ``extra physics'' in fitting baryon oscillations,
S. Habib and K. Heitmann for discussions on initial conditions generation
and Chris Blake, Eric Linder and Ryan Scranton for helpful comments on an
earlier draft.
The simulations were performed on the IBM-SP at NERSC.
MJW was supported in part by NASA and the NSF.

\end{document}